\documentclass[a4paper,12pt]{article}
\usepackage{amssymb}
\usepackage[american]{babel}
\usepackage{colordvi}

\newcommand{\Eq}[1]{Eq. (\ref{#1})}
\newcommand{\Aut}{\mathrm{Aut}}
\newcommand{\Eig}{\mathrm{Eig}}
\newcommand{\folg}[3]{#1_{#2}, \ldots, #1_{#3}}
\newcommand{\id}{\mathrm{id}}

\begin{document}

\bibliographystyle{plain}

\title{\bf Complementarity in \\
classical dynamical systems}
\author{Peter beim Graben \\ 
Institut f\"ur Linguistik, \\
Institut f\"ur Physik, \\
Universit\"at Potsdam, Germany \\
\and
Harald Atmanspacher \\
Institut f\"ur Grenzgebiete der Psychologie und Psychohygiene, \\
Freiburg i.~Br., Germany 
}
\date{}
\maketitle

\bigskip
\centerline{\it Foundations of Physics}
\centerline{{\bf 36}, 291--306 (2006)}

\bigskip

%

\bigskip

\begin{abstract}
The concept of complementarity, originally defined for non-commu\-ting observables of quantum systems with
states of non-vanishing dispersion, is extended to classical dynamical systems with a partitioned phase
space. Interpreting partitions in terms of ensembles of epistemic states (symbols) with corresponding
classical observables, it is shown that such observables are complementary to each other with respect to
particular partitions unless those partitions are generating. This explains why symbolic descriptions
based on an \emph{ad hoc} partition of an underlying phase space description should generally be 
expected to be incompatible. Related approaches with different background and different objectives are discussed.
\end{abstract}

\vfill\eject
\section{Introduction}
\label{intro}
In an algebraic framework \cite{Sakai71, Haag92, AlickiAndriesEA96}, a dynamical system $\Sigma$ (either
quantal or classical) is described by a $\mathrm{C}^*$-algebra of intrinsic observables $\mathfrak{A}$, a
space of positive, linear, normalized functionals $\mathfrak{S} \subset \mathfrak{A}^*$, called \emph{state
space} and a representation $\alpha : G \to \Aut(\mathfrak{A})$ from a group (or a monoid) $G$ into the
automorphism group of $\mathfrak{A}$ mediating the dynamics; i.e. $\Sigma = (\mathfrak{A}, \mathfrak{S},
\alpha, G)$. The distinction between quantum and classical systems relies upon the structure of the algebra
of observables $\mathfrak{A}$. If $\mathfrak{A}$ is commutative, $\Sigma$ is classical, otherwise $\Sigma$ is
quantal. A close correspondence between quantum systems and classical dynamical systems has been revealed by
Koopman and von Neumann in the early 1930s \cite{Koopman31, KoopmanVonNeumann32}. Later, several authors
\cite{AlickiAndriesEA96, AntoniouTasaki93, Alanson92, Kowalski92, LasotaMackey85} further investigated
numerous aspects of this relationship.

Since non-commuting observables, $A, B \in \mathfrak{A}, \: AB \ne BA$, also called \emph{incompatible}
or \emph{complementary} observables, characterize quantum systems, the concept of complementarity appears to
be \emph{prima facie} inapplicable to classical dynamical systems. However, it has been shown that particular
observables can be complementary, i.e.~non-commuting, even in classical stochastic \cite{Tjostheim76,
GustafsonMisra76} and classical dynamical systems \cite{Misra78, AtmanspacherScheingraber87}. For instance,
it has been demonstrated in \cite{AtmanspacherScheingraber87} that in a coarse-grained description of a
chaotic flow there is a pair of operators, the Liouville operator and the information operator, whose
commutator is proportional to the Kolmogorov-Sinai entropy of the flow and does not vanish.

In this paper we extend this result in the spirit of a generalized conception of complementarity which has
been formally axiomatized recently \cite{Atm02} and is not restricted to conventional quantum mechanics. We
consider pairs of arbitrary observables of a classical dynamical system $\Sigma$ and show that those pairs of
observables are \emph{in general} complementary with respect to the coarse-grainings, or partitions, of the
phase space of $\Sigma$. Based on this result, we are able to specify under which conditions partitions can
be called compatible, incompatible, or even complementary.

We start with a brief overview of an algebraic framework for dynamical systems in Sec. \ref{algebra} and
introduce the notion of a phase space partition in Sec. \ref{partisydy}. Using partitions, one can define
suitable symbols by epistemic observables of $\Sigma$ and derive a symbolic dynamics from the original
dynamics in the unpartitioned phase space. In Sec. \ref{epistemacces} the notion of epistemic accessibility
is introduced to distinguish particular partitions by their relation to pointwise states in the original phase space. Section \ref{compobservqm} briefly recalls standard features of compatibility and complementarity in conventional quantum mechanics in order to prepare their application to coarse-grained classical systems in Sec.~\ref{compobservparti}. There we show that only generating partitions give rise to compatible observables, whereas arbitrary partitions generally lead to observables which are incompatible or
complementary. Section~\ref{example} gives a simple example and Sec.~\ref{discu} discusses related approaches, developed for different objectives and from different backgrounds. Section~\ref{summ} summarizes the main 
results.

\section{Algebraic Framework}
\label{algebra}
How is a classical dynamical system $\Sigma$ described within algebraic quantum theory? Because its
observable algebra $\mathfrak{A}$ is commutative, it is isomorphic to a $\mathrm{C}^*$-algebra of complex
valued, continuous functions, $C(X)$, over a locally compact space $X$, which is the spectrum of
$\mathfrak{A}$ \cite[Theorem 1.2.1.]{Sakai71}. One can identify $X$ with the \emph{phase space} of the system
and the physically meaningful observables with the self-adjoined, i.e. real-valued functions $f: X \to
\mathbb{R}$. Since $X$ is a topological space, it is endowed with a Borel $\sigma$-algebra $\mathcal{B} =
\mathcal{B}(X)$. The states are represented by the positive, normalized, linear functionals $\rho: C(X) \to
\mathbb{C}$. They can be related to probability measures $\mu_\rho: \mathcal{B} \to [0, 1]$ by virtue of the
Riesz representation theorem \cite[Theorem 6.3.]{Walters81}, yielding $\rho(f) = \int_X f(x) \, d\mu_\rho(x)$
for all $f \in C(X)$.

Let $M(X, \mathcal{B})$ be the set of measures over $X$ with $\sigma$-algebra $\mathcal{B}$. Then we find, as
a special case, among the measures of $M(X, \mathcal{B})$ the Dirac measures $\delta_x: \mathcal{B} \to \{0,
1\}$, where $\delta_x(A) = 1(0)$, if $x \in(\notin) A$, $A \in \mathcal{B}$ \cite[p. 146]{Walters81}. Thus we
can relate the points $x \in X$ to the measures $\delta_x$ and, hence, to state functionals $\rho_x \in
C(X)^*$. Dirac measures in $M(X,\cal B)$ correspond to pure states in $X$ while all other measures describe
mixtures or statistical states. For time discrete, invertible dynamical systems (to which we
deliberately restrict ourselves in this paper) the group $G = \mathbb{Z}$ of integers is represented by
the iterates of a homeomorphic \emph{flow} map $\Phi: X \to X$ such that $\alpha_t: f \mapsto f \circ \Phi^t$
is an algebra automorphism for all $t \in \mathbb{Z}$. Eventually, we arrive at the well-known description
$\Sigma = ((X, M(X, \mathcal{B})), C(X), \Phi)$.

So far we have discussed the intrinsic structure of $\Sigma$. Using the mathematical apparatus of so-called
GNS-representations \cite{Haag92, Primas90b, AtmanspacherPrimas03}, it is possible to construct a contextual
description of $\Sigma$ from its intrinsic structure. Introducing a reference state $\hat{\rho}$, the
elements of the algebra $\mathfrak{A}$ are represented by bounded operators $\pi_{\hat{\rho}}(A)$
acting on a Hilbert space $\mathfrak{H}_{\hat{\rho}}$. The vectors $\psi_\eta \in \mathfrak{H}_{\hat{\rho}}$
are then identified with the states $\eta \in \mathfrak{S}$ by $\eta(A) = \langle \psi_\eta |
\pi_{\hat{\rho}}(A) \psi_\eta \rangle$, for all $A \in \mathfrak{A}$. In a last step, the weak *-closure
$\pi_{\hat{\rho}}(\mathfrak{A})''$ yields a larger $\mathrm{W}^*$-algebra of \emph{epistemic observables}. At
this level, it is even possible to derive classical observables from a quantum mechanical description
(e.g.~thermodynamic temperature) as emergent properties \cite{Primas90b, Takesaki70}.

Employing the GNS-construction for a classical dynamical system using a reference state given by a measure
$\hat{\mu} \in M(X, \mathcal{B})$ enlarges the $\mathrm{C}^*$-algebra of observables from the continuous
ones, $C(X)$, to the $\mathrm{W}^*$-algebra of $\mathcal{B}$-measurable, $\hat{\mu}$-essentially
bounded functions, $L^\infty(X, \mathcal{B}, \hat{\mu})$ \cite{AlickiAndriesEA96, Primas90b}. In $L^\infty(X,
\mathcal{B}, \hat{\mu})$ one can find discontinuous observables $f$ that are constant over whole regions
$A_i \subset X$ in phase space. Then, by measuring $f$ one observes $f(x) = f(y)$ for all $x, y \in A_i$; 
i.e.~the states $x, y$ are not distinguishable by means of $f$. In this case, we call them \emph{epistemically 
equivalent} with respect to $f$ (in short: $f$-e.e.; \cite{Jauch64, Emch64, WestmorelandSchumacher93}). 
The equivalence classes of $f$-e.e.~states partition the phase space $X$.

\section{Partitions and Symbolic Dynamics}
\label{partisydy}
We are interested in epistemic observables leading to finite partitions of $X$. Let $\mathcal{P}_f$ be a
finite partition of $X$ into classes $\folg A 1 I$ ($I \in \mathbb{N}$) induced by some $f \in L^\infty(X,
\mathcal{B}, \hat{\mu})$. Then we find a ``spectral representation'' of $f$ by
\begin{equation}
\label{simplefunc}
f(x) = \sum_{i = 1}^I f_i \, \chi_{A_i}(x)
\end{equation}
for $x \in X$. The characteristic functions $\chi_{A_i}$ play the role of ``projection operators'' for the
classical system $\Sigma$. The ``spectrum'' of $f$ contains the values $f_i$ assumed by $f$ at the cells of
the partition. Functions given by \Eq{simplefunc} are called \emph{simple functions} \cite{Walters81}. Each
partition defines a host of simple functions by \Eq{simplefunc}. Due to \cite[p. 76]{Walters81} the
collection of sets from $\mathcal{B}$, which are unions of sets of a partition $\mathcal{P}$, is a finite
sub-$\sigma$-algebra of $\mathcal{B}$, called the \emph{partition algebra} $\mathcal{A}(\mathcal{P})$. Let
$\mathcal{P}_1, \mathcal{P}_2$ be finite partitions of $X$. Their \emph{join} is the partition $\mathcal{P}_1
\vee \mathcal{P}_2 =  \{ A_i \cap B_j | A_i \in \mathcal{P}_1, B_j \in \mathcal{P}_2 \}$. Correspondingly,
the joined partition algebra $\mathcal{A}(\mathcal{P}_1) \vee \mathcal{A}(\mathcal{P}_2)$ is the smallest
sub-$\sigma$-algebra of $\mathcal{B}$ containing $\mathcal{A}(\mathcal{P}_1)$ and
$\mathcal{A}(\mathcal{P}_2)$ \cite[Definition 4.3.]{Walters81}.

The value $f(x)$ can be determined by measuring $f$ in state $x$. Continuous measurements can be
described by the time evolution map $\Phi$. Let $x_0 \in X$ be an initial state at time $t=0$. Measuring $f$
yields $x_0 \in A_{i_0}$ for some cell index $1 \le i_0 \le I$ of the partition $\mathcal{P}_f$. After one
time step, the measurement of $f$ yields that the system is in state $\Phi(x_0) \in A_{i_1}$. Thus, we know
that the initial state was in $A_{i_0} \cap \Phi^{-1}(A_{i_1})$, which is a cell of the joined partition
$\mathcal{P}_f \vee \Phi^{-1}(\mathcal{P}_f)$ where $\Phi^{-1}(\mathcal{P})$ is the family of pre-images of
the cells of partition $\mathcal{P}$. Longer lasting measurements of $f$ yield a \emph{refinement}
$\mathcal{P}_f^n = \bigvee_{i = 0}^{n - 1} \Phi^{-i}(\mathcal{P}_f)$ of the original partition. Its partition
algebra is denoted as $\mathcal{A}(\mathcal{P}_f^n) \subseteq \mathcal{B}$. Eventually, we shall also allow
for ``measurements backward in time'' (for $m<0$) leading to refined partitions $^m\mathcal{P}_f^n =
\bigvee_{i = m}^{n - 1} \Phi^{-i}(\mathcal{P}_f)$ and their algebras $\mathcal{A}(^m\mathcal{P}_f^n)
\subseteq \mathcal{B}$.

Fixing the sequence of indices of cells that are visited by a trajectory of the system, one obtains a
sequence $s = \dots a_{i_{-1}} a_{i_0} a_{i_1} a_{i_2}\dots$ of symbols $a_{i_k} \in \mathbf{A}_f$ where the
index set of the partition is interpreted as a finite alphabet $\mathbf{A}_f$ of cardinality $I$. The state
$x_0$ is thereby mapped onto a bi-infinite string $s \in \mathbf{A}_f^\mathbb{Z}$; i.e.~one has constructed a
map $\pi_f: X \to \mathbf{A}_f^\mathbb{Z}$ by the rule $\pi_f(x_0) = s$, iff $\Phi^t(x_0) \in A_{i_t}$, $t
\in \mathbb{Z}$. Although the map $\pi_f$ is usually not invertible, $\pi_f^{-1}$ is applicable in the
set-theoretic sense at subsets of $\mathbf{A}_f^\mathbb{Z}$, e.g.~at strings of finite length. In order to
find their pre-images in $X$, one defines the so-called cylinder sets as sets of bi-infinite sequences with a
common building block beginning with a particular index \cite{CrutchfieldPackard83, McMillan53}. For $t \in
{\mathbb{Z}}, n \in {\mathbb{N}}$ and $\folg a {i_0} {i_{n-1}} \in \mathbf{A}_f$ an \emph{$n$-cylinder} at
time $t$ is given by
\begin{equation}
\label{cylinder}
[a_{i_0}, \dots,  a_{i_{n-1}} ]_t
= \{ s \in \mathbf{A}_f^{\mathbb{Z}} \,|\, s_{t + k} = a_{i_k} ,
\quad k = 0, \dots, n - 1 \} \:.
\end{equation}
The pre-image of a cylinder is $\pi_f^{-1}([\folg a {i_0} {i_{n-1}} ]_t) = \bigcap_{k=0}^{n - 1}
\Phi^{-t-k}(A_{i_k})$ and belongs to $^{t}\mathcal{P}_f^n$. It follows from the definition of $\pi_f$ that
there is a map $\sigma_f : \mathbf{A}_f^\mathbb{Z} \to \mathbf{A}_f^\mathbb{Z}$ shifting all symbols of a
string $s$ one place to the left, such that $\pi_f \circ \Phi = \sigma_f \circ \pi_f$. This \emph{shift map}
defines a \emph{symbolic dynamics} \cite{LindMarcus95}.

Symbolic dynamics provides important qualitative insights into nonlinear dynamical systems and has been
applied for analyzing empirical data (cf.~\cite{beimGrabenSaddyEA00, DawFinneyTracy03, Keller04}). However,
the choice of a phase space partition is somewhat arbitrary; it means to choose a particular epistemic
measurement setup. In this context, questions arise about preferred partitions and invariants.

Particular dynamical systems possess \emph{generators} (\emph{generating partitions}) $\hat\mathcal{P}$
defined by $\mathcal{A}(^{-\infty}\hat\mathcal{P}^\infty) = \mathcal{B} \pmod{0}$, i.e., the backward- and
forward-time iterates of $\hat\mathcal{P}$ approximate the original $\sigma$-algebra of measurable sets
\cite[p. 96]{Walters81}. Generators are known only for very few invertible systems such as the torus map, the
standard map, or the Henon map \cite{AdlerWeiss67, ChristiansenPoliti95, GrassbergerKantz85}. Therefore some
authors suggested algorithms to extract generators from experimental time series \cite{CrutchfieldPackard83,
SteuerMolgedeyEbelingEA01, DavidchackLaiEA00, KennelBuhl03}. An important property of generators is that the
limit of pre-images of the cylinder sets for increasing lengths, $\pi_f^{-1}([\folg a {i_1} {i_\infty}
]_{-\infty}) = \pi_f^{-1}(s)$, where $s = \dots a_{i_{-1}} a_{i_0} a_{i_1} a_{i_2}\dots$ is a bi-infinite
sequence of symbols $a_{i_k} \in \mathbf{A}_f$, contains no more than one point in $X$ \cite[p.
137]{Walters81} (cf. also \cite[p. 203]{LindMarcus95}). In this case, the map $\pi_f$ is a homeomorphism and
the phase space dynamics is topologically equivalent with its coarse-grained symbolic description. If, on the
other hand, $\pi_f^{-1}(s)$ is a non-singleton set $R$ rather than a point, the partition is in a way
\emph{misplaced} and cannot be a generator \cite{BolltStanfordEA2000, BolltStanfordEA01}.

\section{Epistemic Accessibility}
\label{epistemacces}
``Misplaced'' partitions are detrimental for assessing invariants such as the Kol\-mo\-gorov-Sinai entropy
from empirical data, since these invariants will be systematically underestimated \cite{CrutchfieldPackard83,
SteuerMolgedeyEbelingEA01, BolltStanfordEA2000, BolltStanfordEA01}. However, as we shall demonstrate in the
following, such partitions are interesting from another point of view: they correspond to epistemic
observables representing complementary properties of classical dynamical systems.

For this purpose, let us first define a point $x \in X$ as epistemically inaccessible whenever there is no $s
\in \mathbf{A}_f^\mathbb{Z}$ with $\pi_f^{-1}(s) = \{ x \}$. This means that is is not possible, even by
continuous measurement of infinite duration, to tell whether the system is exactly in state $x$ in phase space. The notion of \emph{epistemic accessibility} can easily be introduced in terms of measure theory. 
Suppose $\mathcal{A}(^{-\infty}\mathcal{P}_f^\infty) = \mathcal{B}_f$ is a proper sub-$\sigma$-algebra of 
$\mathcal{B}$ induced by the epistemic observable $f$. Let $\rho \in \mathfrak{S}$ be a state with p.d.f. 
$p \in L^1(X, \mathcal{B}, \hat{\mu})$, where $p \ge 0$ is given with respect to the reference measure 
$\hat{\mu}$ by virtue of the Radon-Nikodym-theorem. We call $\rho$ epistemically accessible with respect to 
$f$ (or $f$-e.a.) if its density $p$ is $\mathcal{B}_f$-measurable. (Recall that a function $p: X \to \mathbb{R}$ is measurable with respect to a $\sigma$-algebra $\mathcal{B}$ if the pre-images of all measurable sets in $\mathbb{R}$ under $p$ are measurable in $\mathcal{B}$.)

Of special interest among the many possible states are those which are described by uniform 
p.d.f.s over some region of phase space $Y \in \mathcal{B}$ with $\hat{\mu}(Y) > 0$
\begin{equation}
\label{unipdf} p(x) = \frac{1}{\hat{\mu}(Y)} \chi_Y(x) \,.
\end{equation}
The associated state $\rho_p \in \mathfrak{S}$ is obviously $f$-e.a. if $Y \in \mathcal{B}_f$. 
Such states define the well-known microcanonical ensembles of thermodynamics \cite{Emch64, beimGraben04}.

\section{Complementary Observables \\ in Conventional Quantum Mechanics}
\label{compobservqm}
As indicated in the Introduction, observables $A, B$ are called \emph{compatible} in the algebraic approach if 
they commute: $[A, B] = 0$ \cite{Atm02}. This definition can be related to the notion of states within 
the conventional Hilbert space representation of quantum mechanics: compatible observables are 
simultaneously diagonalizable, i.e.~there exists a complete orthonormal system of {\it common eigenvectors} 
of $A$ and $B$. Subsequently (Sec.~\ref{compobservparti}) 
we shall demonstrate that a corresponding formulation of compatibility is applicable to classical state spaces that 
are epistemically restricted by partitions.

To illustrate the compatibility of observables in the framework of Hilbert space quantum mechanics, 
let $A$ and $B$ be self-adjoint operators with discrete spectrum acting on state vectors $|\psi\rangle, |\phi\rangle \in \mathfrak{H}$. Since $A, B$ are self-adjoint, their
eigenstates span the whole Hilbert space $\mathfrak{H}$, i.e.~$\mathfrak{H}$ can be decomposed into
direct sums of eigenspaces of $A, B$,
\begin{equation}
\mathfrak{H} = \bigoplus_a \Eig_a(A) \,, \qquad \mathfrak{H} = \bigoplus_b \Eig_b(B)
\end{equation}
where e.g. $\Eig_a(A)$ denotes the eigenspace of $A$ for eigenvalue $a$. The sums run over all different
eigenvalues of $A, B$, respectively.

In order to prove that two observables $A, B$ are \emph{compatible}, iff there is a complete orthonormal 
system of common eigenstates of $A$ and $B$, assume $[A, B] = 0$ and let $|\psi\rangle$ be an eigenvector of 
$A$ with eigenvalue $a$ such that $A |\psi\rangle = a |\psi\rangle$,  
and $B |\psi\rangle = |\phi\rangle$. Then we have $AB |\psi\rangle = A |\phi\rangle$, and 
$BA |\psi\rangle = a B |\psi\rangle = a |\phi\rangle$.
Therefore, by presupposition: $0 = [A, B] |\psi\rangle = (A - a \mathbf{1}) |\phi\rangle$, entailing $A
|\phi\rangle = a |\phi\rangle$, i.e.~$|\phi\rangle \in \Eig_a(A)$. 

In the non-degenerate case all eigenspaces are one-dimensional and we conclude $|\phi\rangle = |\psi\rangle$, i.e.~$|\phi\rangle$ is also an eigenstate of $B$ for eigenvalue 1. In the degenerate case, all eigenspaces of $A$ are invariant under the action of $B$ and \emph{vice versa}. As a consequence, $A$ preserves the direct sum decomposition of $\mathfrak{H}$ into eigenspaces of $B$ and \emph{vice versa}. For the eigenstate
$|\psi\rangle$ of $A$ for eigenvalue $a$ there is a linear combination into basis vectors of $\Eig_a(A)$:
$|\psi\rangle = \sum_i c_i |\psi_i^a\rangle$. This state is taken by $B$ into the eigenstate $|\phi\rangle$
of $A$ for eigenvalue $a$ given by another linear combination with different coefficients $B |\psi\rangle =
\sum_i c_i' |\psi_i^a\rangle$. Now, let $|\chi\rangle$ be an eigenstate of $B$ for eigenvalue $b \ne 0$: $B
|\chi\rangle = b |\chi\rangle$ (note that $|\chi\rangle$ is in general not an eigenstate of $A$). 
For this state we can construct a linear combination of linear combinations
of elements of the eigenspaces of $A$ due to the direct sum decomposition of $\mathfrak{H}$
\begin{equation}
|\chi\rangle = \sum_a \sum_i c_i |\psi_i^a\rangle \,.
\end{equation}
According to the argument above, $B$ acts on this state by
\begin{equation}
B |\chi\rangle = \sum_a \sum_i c_i B |\psi_i^a\rangle = \sum_a \sum_i c_i'  |\psi_i^a\rangle \,.
\end{equation}
On the other hand we know that
\begin{equation}
B |\chi\rangle = b |\chi\rangle = \sum_a \sum_i c_i b |\psi_i^a\rangle \,.
\end{equation}
Therefore, $c_i' = c_i b$, where $i$ enumerates the states of one eigenspace, respectively. Finally, we
obtain $B |\psi_i^a\rangle = b |\psi_i^a\rangle$, i.e.~all base vectors are common eigenstates of 
$A$ and $B$.

For a complete orthonormal system of common eigenstates $|\psi_i\rangle$ of $A$ and $B$ 
the eigenvalue equations read
\begin{eqnarray}
A |\psi_i\rangle &=& a_i |\psi_i\rangle \\
B |\psi_i\rangle &=& b_i |\psi_i\rangle
\end{eqnarray}
From these equations we can deduce $BA |\psi_i\rangle = a_i B |\psi_i\rangle = a_i b_i |\psi_i\rangle$ and
$AB |\psi_i\rangle = b_i A |\psi_i\rangle = a_i b_i |\psi_i\rangle$ and hence $[A, B]| \psi_i\rangle = 0$.
That is, the commutator $[A, B]$ annihilates any common eigenstate $|\psi_i\rangle$. Since these states span
the whole state space, it follows that $[A, B] = 0$. 

Based on this discussion of compatible observables, we call two (self-adjoint) observables $A, B$
\emph{incompatible} if there is no complete orthonormal system of common eigenstates, although there 
may be (some)
common eigenstates. Eventually, we call two observables $A, B$ \emph{complementary} if they have 
no eigenstate in common. If $A$ and $B$ are projectors, the latter definition entails that there is no state 
which makes $A$ and $B$ simultaneously truth-definite \cite{RaggioRieckers83}.

\section{Complementary Observables and Partitions \\ in Classical Systems}
\label{compobservparti}
Given an algebra of observables $\mathfrak{A}$ and a state space $\mathfrak{S}$ of a classical dynamical
system, applying a state functional $\rho \in \mathfrak{S}$ to an observable $A \in \mathfrak{A}$ yields the \emph{expectation value} of $A$ in state $\rho$. We define the \emph{dispersion} of $A$ in $\rho$ by
\begin{equation}
\label{dispersion} \Delta_\rho(A) = \sqrt{\rho(A^2) - \rho(A)^2} \:.
\end{equation}
If $\Delta_{\rho}(A) = 0$, $\rho$ is an \emph{eigenstate} of $A$ with eigenvalue $a = \rho(A)$. A common
eigenstate of two observables $A$ and $B$ is a state $\eta \in \mathfrak{S}$ such that $\Delta_{\eta}(A) =
\Delta_{\eta}(B) = 0$.

Applying the notion of compatibility defined in Sec.~5 to classical systems, we find that 
two observables $A, B \in \mathfrak{A}$ are \emph{compatible} insofar as a complete orthonormal 
system of commonly shared eigenstates is given by the Dirac distributions 
$\rho(A) = \int \delta(x - x_0) A(x) d\hat{\mu}(x)$ with
respect to pure states $x_0$ which are evaluated pointwise in phase space, i.e.~they are
\emph{dispersion-free}, $\Delta_{\rho}(A) = 0$ for all $A \in \mathfrak{A}$ \cite{Misra67}. However, pure
classical states can be epistemically inaccessible with respect to a given partition. In this case we
call two observables $A, B \in \mathfrak{A}$ \emph{incompatible} if there is no complete orthonormal system
of common epistemically accessible eigenstates. Accordingly, $A$ and $B$ are 
\emph{maximally incompatible}, or \emph{complementary}, if they have no epistemically accessible
eigenstate in common.

Then, the important question arises whether statistical states can be eigenstates of classical observables. Let $\rho_p \in \mathfrak{S}$ be a statistical state with p.d.f. $p$, and let $h \in \mathfrak{A}$ be a classical observable, i.e.~a measurable function $h:X \to \mathbb{R}$. Collecting all $x \in X$ with $h(x) = 1$ yields a set of states $Y = h^{-1}(\{1\}) \subset X$. If $\int_Y p(x) d\hat{\mu}(x) > 0$, we can construct another state $\rho'$ with density $p'$ according to
\begin{equation}
\label{stateigenst}
p'(x) = \frac{1}{\int_Y p(x) d\hat{\mu}(x)} p(x) \chi_Y(x) \,.
\end{equation}
Then $\rho'$ is an eigenstate of $h$. On the other hand, we can easily find an observable $h$ that has the
microcanonical ensemble \Eq{unipdf} as its eigenstate, namely by setting
\begin{equation}
\label{unieigst}
h(x) = \left\{ \begin{array}{r@{\quad:\quad}l}
\mbox{constant} & x \in Y \\
\mbox{arbitrary} & x \notin Y \,.
\end{array}
\right.
\end{equation}

Now the crucial point is whether or not a statistical eigenstate of a classical observable is epistemically
accessible. If states are epistemically inaccessible, then even classical observables can be incompatible or complementary. A corresponding scenario has recently been suggested based on a toy model in
\cite{Spekkens04}. 

Complementary observables in a coarse-grained, i.e.~partitioned, phase space of a classical dynamical system arise in the following way. Let $\mathcal{P}_f$ be a non-generating partition of the phase space $X$ induced by an epistemic observable $f$ and $\mathcal{B}_f \subset \mathcal{B}$ the associated sub-$\sigma$-algebra. Then there exist non-singleton sets $R_s = \pi_f^{-1}(s) \in \mathcal{B}_f$ and $R_r = \pi_f^{-1}(r) \in \mathcal{B}_f$ for some bi-infinite sequences $s, r \in \mathbf{A}_f^\mathbb{Z}$ ($s \ne r$) of the symbolic dynamics of the system. If $\hat{\mu}(R_s) > 0$ and $\hat{\mu}(R_r) > 0$, we can define uniform p.d.f.'s~$p_s(x) = \chi_{R_s}(x) / \hat{\mu}(R_s)$, $p_r(x) = \chi_{R_r}(x) / \hat{\mu}(R_r)$ (\Eq{unipdf}), and, hence, statistical states $\rho_{p_s}$, $\rho_{p_r}$ both of which are $f$-e.a.~ Now we define observables
\begin{eqnarray}
\label{clcomp}
h_s(x) &=& \left\{ \begin{array}{r@{\quad:\quad}l}
\mbox{constant} & x \in R_s \\
\mbox{arbitrary but neither 0 nor 1} & x \notin R_s
\end{array}
\right. \\
h_r(x) &=& \left\{ \begin{array}{r@{\quad:\quad}l}
\mbox{constant} & x \in R_r \\
\mbox{arbitrary but neither 0 nor 1} & x \notin R_r
\end{array}
\right.
\end{eqnarray}
according to \Eq{unieigst}, since $\chi_{R_s}$ and $\chi_{R_r}$ vanish outside $R_s$ and $R_r$,
respectively (note that this point was not correctly taken into account in \cite{beimGraben04}). Then
$\rho_{p_s}$ is an eigenstate of $h_s$ and $\rho_{p_r}$ is an eigenstate of $h_r$. However, there is no
common eigenstate of $h_s$ and $h_r$ which is $f$-e.a.~since $R_s \cap R_r = \emptyset$. Hence, the
observables $h_s$ and $h_r$ are complementary.

Consider another epistemic observable $g$ yielding a ``misplaced'' partition $\mathcal{P}_g$ and its set
algebra $\mathcal{B}_g \subset \mathcal{B}$. We call the partitions $\mathcal{P}_f$ and $\mathcal{P}_g$
complementary if no $f$-e.a.~eigenstate of an observable is also a $g$-e.a.~eigenstate of an observable, and
\emph{vice versa}. Using the $\sigma$-algebras $\mathcal{B}_f, \mathcal{B}_g$ of the partitions
$\mathcal{P}_f, \mathcal{P}_g$, respectively, we can conveniently distinguish between compatible,
incompatible and maximally incompatible partitions in the following way. Any two partitions $\mathcal{P}_1,
\mathcal{P}_2$ are \emph{compatible} if their $\sigma$-algebras $\mathcal{B}_1, \mathcal{B}_2$ are identical
up to sets of $\hat{\mu}$-measure zero, $\mathcal{B}_1 = \mathcal{B}_2 \pmod{0}$. They are
\emph{incompatible} if their $\sigma$-algebras are not identical (up to $\hat{\mu}$-zero-sets),
$\mathcal{B}_1 \ne \mathcal{B}_2 \pmod{0}$. They are \emph{maximally incompatible} if their $\sigma$-algebras
are disjoint up to the trivial overlap $\mathcal{B}_1 \cap \mathcal{B}_2 = \{X\}$, i.e. $\mathcal{B}_1 \cap
\mathcal{B}_2 = \emptyset \pmod{1}$, because $\hat{\mu}(X) = 1$.

Compatible partitions $\mathcal{P}_1, \mathcal{P}_2$ have the same pairs of compatible observables as
all states which are $f_1$-e.a. are also $f_2$-e.a (note that one could also construct pairs of incompatible
observables using statistical states as we did above; however it would be impossible constructing pairs of
complementary observables for compatible partitions). For incompatible partitions all kinds of
observables are possible. Only for maximally incompatible partitions all $f_1$-e.a. states are not
$f_2$-e.a. and \emph{vice versa}. This holds in particular for eigenstates of observables. Therefore, all
observables that have eigenstates are complementary to each other with respect to maximally incompatible
partitions. Hence, such partitions can be called complementary.

Whenever $\mathcal{P}_1$ and $\mathcal{P}_2$ are generators, their algebras obey $\mathcal{B}_1 = \mathcal{B} = \mathcal{B}_2 \pmod{0}$: generators are always compatible. On the other hand, two incompatible or complementary partitions cannot both be generating and, thus, do not lead to topologically equivalent
symbolic dynamics. Moreover, a symbolic dynamics based on a non-generating partition is not topologically
equivalent with the original phase space dynamics.

\section{A Simple Example}
\label{example}
In order to illustrate this argumentation, let us consider a simple example motivated by the torus map
\cite{AdlerWeiss67}. The phase space of this system is the 2-torus which can be uniquely mapped onto the unit
square $[0, 1]^2$ with periodic boundaries. We shall consider the identity function $\Phi \equiv \id: [0,
1]^2 \to [0, 1]^2$ and a binary partition $\mathcal{P}_1$ of $X = [0, 1]^2$ given by $A_0 = \{ (x_1, x_2) \in
[0, 1]^2 | x_1 \le 0.5 \}$, $A_1 = \{ (x_1, x_2) \in [0, 1]^2 | x_1 > 0.5 \}$. The symbolic dynamics
corresponding to this non-generating partition has only two fixed points $s = \dots 00.00 \dots =
{}^{-\infty}0.0^\infty$ and $r = {}^{-\infty}1.1^\infty$ where the decimal point indicates the observation
time. The pre-images of these sequences are simply $R_s = \pi^{-1}(s) = A_0$ and $R_r = \pi^{-1}(r) = A_1$
with $R_s \cap R_r = \emptyset$. The sets $R_s, R_r$ support two uniformly distributed p.d.f.'s which
correspond to two eigenstates $\rho_s, \rho_r$ of observables $h_s, h_r$ assuming constant values at
$R_s, R_r$, respectively. Hence, $h_s$ and $h_r$ are complementary with respect to $\mathcal{P}_1$.

Consider the partition $\mathcal{P}_2 = \{ B_0, B_1 \}$ of the unit square given as $B_0 = \{ (x_1, x_2) \in [0, 1]^2 | x_2 \le 0.5 \}$, $B_1 = \{ (x_1, x_2) \in [0, 1]^2 | x_2 > 0.5 \}$ in addition to $\mathcal{P}_1$ in the above example. As for $\mathcal{P}_1$, $\mathcal{P}_2$ defines complementary observables whose eigenstates are supported by $B_0, B_1$. Moreover, the intersections $A_i \cap B_j$ are not epistemically accessible with respect to both partitions. Therefore, $\mathcal{P}_1$ and $\mathcal{P}_2$ are complementary as well.

\section{Related Approaches}
\label{discu}
Our approach resembles other kinds of ``epistemic equivalence'' that have been used to describe 
physical measurement. Emch and Jauch introduced the concept of \emph{macroscopic equivalence} 
of quantum states to deal with the quantum mechanical measurement problem \cite{Jauch64, Emch64}. They defined two states (density matrices) as macroscopically equivalent if their expectation values are the same in the sense that they are indistinguishable by means of the measuring device. 

Jauch \cite{Jauch64} proved that equivalence classes of such states, called \emph{macro\-states}, are stable under the action of classical, i.e.~compatible, observables. He also proved that the particular pure state described by a Schr\"odinger equation of the measurement device on the one hand, and the particular mixture resulting from the Heisenberg cut between the quantum object and its environment on the other hand, belong to the same macrostate and are therefore epistemically indistinguishable provided that the measurement device is described as a classical apparatus \cite{Bohr32}. 

Emch \cite{Emch64}, by contrast, defined a macrostate as a linear combination of microcanonical ensembles. More important, however, for the purpose of the present paper is the \emph{coarse-graining} that he applied to the algebra of observables. To each (microscopic) observable he assigned its macroscopic equivalent, a linear combination of projectors whose coefficients are the expectation values of the microscopic observable in the microcanonical states. This construction is analogous to our exposition of phase space partitions as classes of simple functions in Sect.~\ref{partisydy}.

A rather different approach is due to Westmoreland and Schumacher \cite{WestmorelandSchumacher93}, who used topological criteria for the definition of equivalence by physical measurement. They argued that classical observables as continuous functions in a proper phase space are not precisely measurable. Given some open interval for the measurement error, the pre-image of this interval under an observable is an open set in phase space. Representing propositions by subsets of the phase space, two propositions are physically equivalent if either their topological closures or their interiors are equal. 

Westmoreland and Schumacher showed that these equivalence relations lead to non-Boolean logics even for classical dynamical systems. However, these logics do not coincide with non-distributive quantum logics. 
Moreover, only measurements in finite time were addressed. Our approach, by contrast, includes the option of arbitrarily high precision by continuous measurements of increasing measuring time. This is possible when a finite partition of the phase space providing the propositions of the logical calculus is generating. For
non-generating partitions, quantum-like features can arise even for classical systems.

Generating partitions yield symbolic representations that are topologically equivalent to the original
phase space dynamics. This \emph{robustness} or \emph{stability} criterion can be used as a 
constraint for the construction of proper symbolic descriptions. While the choice of symbolic 
descriptions based on generators is essentially motivated  by stability issues, a viable alternative 
relies basically on information theoretical ideas. This alternative is embedded in the framework of 
computational mechanics, as pioneered by Crutchfield and coworkers (see \cite{Shalizi01} for a 
comprehensive review). A recent paper by Shalizi and Moore \cite{Shalizi05}    
is useful to clarify the relation between their and our approach.

A key notion in computational mechanics is the notion of a ``causal state''. Its definition is based on the equivalence class of histories of a process that are equivalent for predicting the future of the process (see \cite{Israeli04} for a related idea applied to cellular automata). Since any prediction method induces a partition of the phase space of the system, the choice of an appropriate partition is crucial. If the partition is too fine, too many (irrelevant) details of the process are taken into account; if the partition is too coarse, not enough (relevant) details are considered.

As described in detail in \cite{Shalizi05}, it is possible to determine partitions leading to causal states. This is achieved by minimizing their statistical complexity, the amount of information which the partition encodes about the past. The procedure starts with an initial choice of a partition guided by a context that is assumed to be relevant for a suitable description of the process under study. If this partition does not provide causal states already, it is iteratively refined or coarsened as long as the resulting states are causal in the sense that the space of admissible symbol sequences is partitioned into
classes predicting equivalent futures. 


Causal states depend on the ``subjectively'' chosen initial partition but are then ``objectively'' fixed by the underlying dynamics. This has been expressed succinctly in \cite{Shalizi05}: Nature has no preferred questions, but to any selected question it has a definite answer. Quite similarly, our notion of robust epistemic states combines the ``subjective'' notion 
of coarse-graining with an ``objective''  way to determine proper partitions as they are generated by the underlying dynamics of the system. 

Finally, the notion of incompatible descriptions due to non-generating partitions provides insight concerning particular quantum-like features such as ``Brownian entanglement'' in classical systems as reported recently \cite{Allah05}. Two particles undergoing Brownian motion were shown to create an analogue to quantum entanglement for coarse-grained velocities. From the perspective of our approach, it may be conjectured that this coarse-graining introduces non-generating partitions inducing the correlations responsible for ``Brownian entanglement''. The observation in \cite{Allah05} that these correlations disappear for an increasingly refined resolution of the coarse-graining is consistent with an \emph{asymptotic} epistemic accessibility
of classical pure states.

\section{Summary}
\label{summ}
Measurements of classical systems are generally ambiguous concerning an individual, pointwise representation
of the state of the system in phase space. This ambiguity can be expressed by means of a coarse-graining
induced by classes of epistemically equivalent states partitioning the phase space and introducing an
``epistemic quantization''. Rarely such partitions are generators allowing us to approximate the individual
states by continuous measurements leading to a dynamical refinement of the partition. Only in this
\emph{special} case will individual states be epistemically accessible in the limit of infinite forward
observation time and knowledge of the infinite past.

The \emph{general} situation of ``misplaced'' or non-generating partitions is interesting for other reasons,
however. In an algebraic framework we showed that classical observables can be complementary with respect to
particular partitions if not all states in the full state space are epistemically accessible. In this case,
different partitions can themselves be considered as complementary if their Borel $\sigma$-algebras are
disjoint up to their trivial overlap. Hence, observables and partitions associated with epistemic states can
be complementary even if their underlying description is purely classical.

Related ideas have been discussed in the literature since the 1960s. Emch \cite{Emch64} and Jauch
\cite{Jauch64} used the notion of epistemic indistinguishability of ontic states within so-called
macrostates to describe measurement. Shalizi and Moore \cite{Shalizi05} developed an information 
theoretical approach to construct descriptions of systems with robust epistemic states providing Markov processes. Westmoreland and Schumacher \cite{WestmorelandSchumacher93} showed that classical systems can obey a non-Boolean logic which, however, is not identical with non-Boolean quantum logic. Recent observations of quantum-like behavior in classical systems by Allahverdyan \emph{et al.} \cite{Allah05} can be interpreted due to an incompatibility of non-generating partitions created by an \emph{ad hoc} coarse-graining.

Since Borel algebras induced by epistemic observables define \emph{contextual topologies}, we expect that complementary partitions will be significant for issues of reduction and emergence in hierarchical descriptions of complex systems \cite{Primas77, Primas98} even beyond physics.

\section*{Acknowledgments}
We gratefully acknowledge useful discussions with Carsten Allefeld, Guido Bacciagaluppi, 
Werner Ehm and Hans Primas, and helpful comments by two anonymous referees. 
This work has been supported by the Deutsche Forschungsgemeinschaft within the research group
on ``Conflicting Rules in Cognitive Systems''.


\begin{thebibliography}{10}

\bibitem{AdlerWeiss67}
R.~L. Adler and B.~Weiss.
\newblock Entropy, a complete metric invariant for automorphisms of the torus.
\newblock {\em Proc. Natl. Acad. Sci. U.S.A.}, 57:1573 -- 1576, 1967.

\bibitem{Alanson92}
T.~Alanson.
\newblock A ``quantal'' {H}ilbert space formulation for nonlinear dynamical
  systems in terms of probability amplitudes.
\newblock {\em Phys. Lett.}, A 163:41 -- 43, 1992.

\bibitem{AlickiAndriesEA96}
R.~Alicki, J.~Andries, M.~Fannes, and P.~Tuyls.
\newblock An algebraic approach to the {K}olmogorov-{S}inai entropy.
\newblock {\em Rev. Math. Phys.}, 8(2):167 -- 184, 1996.

\bibitem{Allah05}
A.E.~Allahverdyan, A.~Khrennikov, and Th.M.~Nieuwenhuizen.
\newblock Brownian entanglement.
\newblock manuscript, arXiv: quant-ph/0412132. 
\newblock Accepted for publication in Physical Review A.

\bibitem{AntoniouTasaki93}
I.~Antoniou and S.~Tasaki.
\newblock Generalized spectral decomposition of mixing dynamical systems.
\newblock {\em Int. J. Quant. Chem.}, 46:425 -- 474, 1993.

\bibitem{AtmanspacherPrimas03}
H.~Atmanspacher and H.~Primas.
\newblock Epistemic and ontic quantum realities.
\newblock In L.~Castell and O.~Ischebeck, editors, {\em Time, Quantum, and
  Information}, pages 301 -- 321. Springer, Berlin, 2003.

\bibitem{Atm02}
H.~Atmanspacher, H.~R\"omer, and H.~Walach.
\newblock Weak quantum theory: Complementarity and entanglement in physics and
beyond.
\newblock {\em Foundations of Physics}, 32:379--406, 2002.

\bibitem{AtmanspacherScheingraber87}
H.~Atmanspacher and H.~Scheingraber.
\newblock A fundamental link between system theory and statistical mechanics.
\newblock {\em Foundations of Physics}, 17(9):939 -- 963, 1987.

\bibitem{beimGraben04}
P.~beim Graben.
\newblock Incompatible implementations of physical symbol systems.
\newblock {\em Mind and Matter}, 2(2):29--51, 2004.

\bibitem{beimGrabenSaddyEA00}
P.~beim Graben, J.~D. Saddy, M.~Schlesewsky, and J.~Kurths.
\newblock Symbolic dynamics of event--related brain potentials.
\newblock {\em Phys. Rev. E}, 62(4):5518 -- 5541, 2000.

\bibitem{Bohr32}
N.~Bohr.
\newblock Chemistry and the quantum theory of atomic constitution.
\newblock {\em J. Chem. Soc.}, 134:349 -- 384, 1932.

\bibitem{BolltStanfordEA2000}
E.~M. Bollt, T.~Stanford, Y.-C. Lai, and K.~\.{Z}yczkowski.
\newblock Validity of threshold-crossing analysis of symbolic dynamics from
  chaotic time series.
\newblock {\em Phys. Rev. Lett.}, 85(16):3524 -- 3527, 2000.

\bibitem{BolltStanfordEA01}
E.~M. Bollt, T.~Stanford, Y.-C. Lai, and K.~\.{Z}yczkowski.
\newblock What symbolic dynamics do we get with a misplaced partition? {O}n the
  validity of threshold crossings analysis of chaotic time-series.
\newblock {\em Physica D}, 154:259 -- 286, 2001.

\bibitem{ChristiansenPoliti95}
F.~Christiansen and A.~Politi.
\newblock Generating partition of the standard map.
\newblock {\em Phys. Rev. E}, 51(5):3811 -- 3814, 1995.

\bibitem{CrutchfieldPackard83}
J.~P. Crutchfield and N.~H. Packard.
\newblock Symbolic dynamics of noisy chaos.
\newblock {\em Physica D}, 7:201 -- 223, 1983.

\bibitem{DavidchackLaiEA00}
R.~L. Davidchack, Y.-C. Lai, E.~M. Bollt, and M.~Dhamala.
\newblock Estimating generating partitions of chaotic systems by unstable
  periodic orbits.
\newblock {\em Phys. Rev. E}, 61(2):1353 -- 1356, 2000.

\bibitem{DawFinneyTracy03}
C.~S. Daw, C.~E.~A. Finney, and E.~R. Tracy.
\newblock A review of symbolic analysis of experimental data.
\newblock {\em Rev. Sci. Instrum.}, 74:915 -- 930, 2003.

\bibitem{Emch64}
G.~Emch.
\newblock Coarse-graining in {L}iouville space and master equation.
\newblock {\em Helv. Phys. Act.}, 37:532 -- 544, 1964.

\bibitem{GrassbergerKantz85}
P.~Grassberger and H.~Kantz.
\newblock Generating partitions for the dissipative {H}enon map.
\newblock {\em Phys. Lett.}, 113A(5):235 -- 238, 1985.

\bibitem{GustafsonMisra76}
K.~Gustafson and B.~Misra.
\newblock Canonical commutation relations of quantum mechanics and stochastic
  regularity.
\newblock {\em Lett. Math. Phys.}, 1:275 -- 280, 1976.

\bibitem{Haag92}
R.~Haag.
\newblock {\em Local Quantum Physics: Fields, Particles, Algebras}.
\newblock Springer, Berlin, 1992.

\bibitem{Israeli04}
N.~Israeli and N.~Goldenfeld.
\newblock On computational irreducibility and the predictability of complex physical systems.
\newblock {\it Phys.~Rev.~Lett.} 92: 074105, 2004.

\bibitem{Jauch64}
J.~M. Jauch.
\newblock The problem of measurement in quantum mechanics.
\newblock {\em Helv. Phys. Act.}, 37:293 -- 316, 1964.

\bibitem{Keller04}
K.~Keller and K.~Wittfeld.
\newblock Distances of time series components by means of symbolic dynamics.
\newblock {\em Int.~J.~Bif.~Chaos} 14: 693--704, 2004.

\bibitem{KennelBuhl03}
M.~B. Kennel and M.~Buhl.
\newblock Estimating good discrete partitions from observed data: Symbolic
  false nearest neighbors.
\newblock {\em Phys. Rev. Lett.}, 91(8):1--4, 2003.

\bibitem{Koopman31}
B.~O. Koopman.
\newblock Hamiltonian sytems and transformations in {H}ilbert space.
\newblock {\em Proc. Natl. Acad. Sci. U.S.A.}, 17:315 -- 318, 1931.

\bibitem{KoopmanVonNeumann32}
B.~O. Koopman and J.~von Neumann.
\newblock Dynamical systems of continuous spectra.
\newblock {\em Proc. Natl. Acad. Sci. U.S.A.}, 18:255 -- 262, 1932.

\bibitem{Kowalski92}
K.~Kowalski.
\newblock Linearization transformations for non-linear dynamical systems:
  {H}ilbert space approach.
\newblock {\em Physica A}, 180:156 -- 170, 1992.

\bibitem{LasotaMackey85}
A.~Lasota and M.C. Mackey.
\newblock {\em Probabilistic Properties of Deterministic Systems}.
\newblock Cambridge University Press, 1985.

\bibitem{LindMarcus95}
D.~Lind and B.~Marcus.
\newblock {\em An Introduction to Symbolic Dynamics and Coding}.
\newblock Cambridge University Press, Cambridge (UK), 1995.

\bibitem{McMillan53}
B.~McMillan.
\newblock The basic theorems of information theory.
\newblock {\em Ann. Math. Statist.}, 24:196 -- 219, 1953.

\bibitem{Misra67}
B.~Misra.
\newblock When can hidden variables be excluded in quantum mechanics.
\newblock {\em Il Nuovo Cimento}, 47(4):841 -- 859, 1967.

\bibitem{Misra78}
B.~Misra.
\newblock Nonequilibrium entropy, {L}yapunov variables, and ergodic properties
  of classical systems.
\newblock {\em Proc. Natl. Acad. Sci. U.S.A.}, 75:1627 -- 1631, 1978.

\bibitem{Primas77}
H.~Primas.
\newblock Theory reduction and non-{B}oolean theories.
\newblock {\em J. Math. Biol.}, 4:281 -- 301, 1977.

\bibitem{Primas90b}
H.~Primas.
\newblock Mathematical and philosophical questions in the theory of open and
  macroscopic quantum systems.
\newblock In A.~I. Miller, editor, {\em Sixty-two Years of Uncertainty:
  Historical, Philosophical and Physics Inquries into the Foundation of Quantum
  Mechanics}, pages 233 -- 257. Plenum Press, New York, 1990.

\bibitem{Primas98}
H.~Primas.
\newblock Emergence in exact natural sciences.
\newblock {\em Acta Polytechnica Scandinavica}, Ma-91:83 -- 98, 1998.

\bibitem{RaggioRieckers83}
G.~A. Raggio and A.~Rieckers.
\newblock Coherence and incompatibility in $\mathrm{W}^*$-algebraic quantum
  theory.
\newblock {\em Int. J. Theor. Phys.}, 22(3):267 -- 291, 1983.

\bibitem{Sakai71}
S.~Sakai.
\newblock {\em $C^*$-Algebras and $W^*$-Algebras}.
\newblock Springer, Berlin, 1971.

\bibitem{Shalizi01}
C.R.~Shalizi and J.P.~Crutchfield. 
\newblock Computational mechanics: pattern and prediction, structure and simplicity.
\newblock {\it J.~Stat.~Phys.} 104:817--879, 2001.

 \bibitem{Shalizi05}
C.R.~Shalizi and C.~Moore. 
\newblock What is a macrostate? Subjective observations and objective dynamics.
\newblock manuscript, arXiv: cond-mat/0303625.

\bibitem{Spekkens04}
R.W.~Spekkens.
\newblock In defense of the epistemic view of quantum states: A toy theory.
\newblock manuscript, arXiv: quant-ph/0401052.

\bibitem{SteuerMolgedeyEbelingEA01}
R.~Steuer, L.~Molgedey, W.~Ebeling, and M.~A. Jim\'{e}nez-Montano.
\newblock Entropy and optimal partition for data analysis.
\newblock {\em Eur. Phys. J. B}, 19:265 -- 269, 2001.

\bibitem{Takesaki70}
M.~Takesaki.
\newblock Disjointness of the {KMS}-states of different temperatures.
\newblock {\em Commun. Math. Phys.}, 17:33 -- 41, 1970.

\bibitem{Tjostheim76}
D.~Tj{\o}stheim.
\newblock A commutation relation for wide sense stationary processes.
\newblock {\em SIAM J. Appl. Math.}, 30(1):115 -- 122, 1976.

\bibitem{Walters81}
P.~Walters.
\newblock {\em Introduction to Ergodic Theory}.
\newblock Springer, Tokyo, 1981.

\bibitem{WestmorelandSchumacher93} M.~D.~Westmoreland and B.~W.~Schumacher.
\newblock Non-{B}oolean derived logics for classical systems.
\newblock {\em Phys. Rev. A}, 48:977 -- 985, 1993.


\end{thebibliography}
\end{document}